\documentclass[twocolumn,preprintnumbers,amsmath,amssymb,prb]{revtex4}

\usepackage{graphicx}
\usepackage{dcolumn}
\usepackage{bm}
\usepackage{amssymb}
\begin{document}

\title{Effect of magnetic field on intersubband polaritons in a quantum well:
Strong to weak coupling conversion}

\author{A. A. Pervishko$^{1,2}$}
\author{O. V. Kibis$^{3,1}$}\email{Oleg.Kibis(c)nstu.ru}
\author{I. A. Shelykh$^{1,2,4}$}

\affiliation{$^1$Division of Physics and Applied Physics, Nanyang
Technological University 637371,  Singapore} \affiliation{$^2$ITMO
University, St. Petersburg 197101, Russia}
\affiliation{$^3$Department of Applied and Theoretical Physics,
Novosibirsk State Technical University, Novosibirsk 630073,
Russia} \affiliation{$^4$Science Institute, University of Iceland
IS-107, Reykjavik, Iceland}

\begin{abstract}
We investigate theoretically the effect of a magnetic field on
intersubband polaritons in an asymmetric quantum well placed
inside an optical resonator. It is demonstrated that the
field-induced diamagnetic shift of electron subbands in the well
increases the broadening of optical lines corresponding to
intersubband electron transitions. As a consequence, the magnetic
field can switch the polariton system from the regime of strong
light-matter coupling to the regime of weak one. This effect paves
a way to the effective control of polaritonic devices with a
magnetic field.
\end{abstract}
\maketitle

In recent decades, the significant technological progress in
fabrication of optical resonators (microcavities) with high
quality factor was achieved. This led to both experimental and
theoretical development of the concept of strong light-matter
coupling, which covers various electron-photon processes
accompanying by the oscillatory energy exchange between an
electromagnetic field confined in a microcavity and electrons in
condensed-matter structures placed inside the microcavity
\cite{Kavokin,KavokinBook,Deveaud,Sanvitto,Weisbuch,Saba,
Kasprzak, Amo,Kibis_11, Liew,Kibis_13, Byrnes}. Currently, the
regime of strong light-matter coupling has been realized
experimentally in a variety of condensed-matter structures,
including quantum wells (QWs) \cite{Christmann,Deng, Egorov},
quantum dots \cite{Reithmaier,Yoshie,Bloch_2005}, quantum wires
\cite{Citric,Wertz,Trichet, Manni, Tanese}, and others.
Particularly, the energy exchange between electrons in QW and a
confined electromagnetic field in a microcavity leads to the
oscillating transitions between different electron subbands in QW.
These periodical optical transitions of electrons can be described
formally as a composite electron-photon quasi-particle called
``intersubband polariton'' \cite{Dini,
Dupont,Todorov,Kyriienko,Zaluzny,KyriienkoSO,Murphy,Zanotto}.
Immediately after experimental discovery of the intersubband
polaritons, they attracted great attention of research community.
This was caused by their unique physical properties which meet the
needs of various modern optoelectronic devices operating in a wide
frequency range from infrared to terahertz domains, including
quantum cascade lasers \cite{Colombelli, Sapienza},  light
emitting devices \cite{ Jouy, Geiser1, Colombelli2015}   and
photodetectors \cite{Liu}. In order to control the devices, the
frequency tuning of intersubband polariton should be elaborated.
Currently, the tuning is realized with the control of light-matter
coupling constant by the electrical gating applied to QW
\cite{Anappara, Geiser}. In the present study, we propose the
alternative method of the frequency tuning by an external magnetic
field. In what follows, we will show theoretically that the
magnetic field can effectively control the intersubband polaritons
by switching the electron-photon system in QW from the regime of
strong light-matter coupling to the weak one.

Generally, the regime of strong light-matter coupling is realized
when the coherent light-matter interaction overcomes the
characteristic damping. In this regime, the energy is periodically
exchanged between the ``light part'' and ``matter part'' of the
light-matter system with the Rabi frequency, $\Omega_R$, which is
given by the expression \cite{Kavokin}
\begin{eqnarray}
\hbar\Omega_R=\sqrt{g_R^2-\Gamma^2}, \label{eq:Om}
\end{eqnarray}
where $g_R$ is the characteristic light-matter coupling constant,
and $\Gamma$ is the characteristic linewidth depending on the
damping in the system. It follows from Eq.~(\ref{eq:Om}) that the
condition of strong light-matter coupling can be written in the
most general form as $g_R>\Gamma$. The opposite case,
$g_R\leq\Gamma$, corresponds to the regime of weak light-matter
coupling, when there is no oscillating processes in the
light-matter system.

\begin{figure}[h!]
\includegraphics[width=0.47\textwidth]{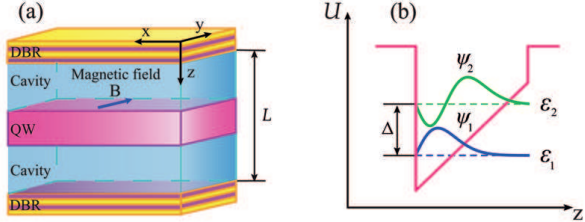}
\caption{(a) Sketch of the system under consideration: (a) the
quantum well (QW) placed inside the planar cavity formed by the
distributed Bragg reflectors (DBRs) in the presence of an in-plane
magnetic field, ${B}$; (b) the asymmetric confining potential of
the QW, $U(z)$, with the edge energies of the two first electron
subbands, $\varepsilon_{1,2}$, and the corresponding electron wave
functions, $\psi_{1,2}(z)$.} \label{fig:fig1}
\end{figure}

\begin{figure}[h!]
\includegraphics[width=0.47\textwidth]{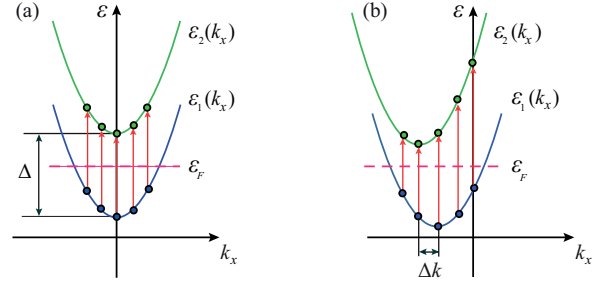}
\caption{Energy spectrum of the two first electron subbands,
$\varepsilon_{1,2}(k_x)$, in the asymmetric QW: (a) in the absence
of a magnetic field; (b) in the presence of an in-plane magnetic
field directed along the $y$ axis. The red arrows mark optical
transitions of electrons, which are induced by the cavity photons
at different electron wave vectors, $k_x$. The electron gas is
assumed to fill electron states in the first subband under the
Fermi energy, $\varepsilon_F$.} \label{fig:fig11}
\end{figure}

Let us apply the aforesaid to analyze electron-photon processes in
a QW placed inside a planar microcavity (see
Fig.~\ref{fig:fig1}a). We will assume the confining potential of
QW, $U(z)$, to be asymmetric (see Fig.~\ref{fig:fig1}b) and
restrict the consideration by two first electron subbands in QW
(see Fig.~\ref{fig:fig11}). In the considered system, the regime
of strong light-matter coupling corresponds to the intersubband
Rabi oscillations of electrons with the frequency (\ref{eq:Om}),
where $\Gamma$ is the characteristic linewidth of the intersubband
transitions induced by cavity photons. Since the characteristic
value of the photon momentum is very small, these optical
transitions can be pictured approximately as vertical (see red
arrows in Fig.~\ref{fig:fig11}). As to the electron-photon
coupling constant, it is
\begin{equation}\label{gR}
g_R=g(q_0)\sqrt{Sn_e},
\end{equation}
where
\begin{eqnarray}
g(q)=\sqrt{\frac{|d_{12}|^2\Delta^2}{S\hbar \epsilon_0\epsilon
L\omega_0(q)}\frac{q^2}{(\pi/L)^2+q^2}} \label{eq:g}
\end{eqnarray}
is the matrix element of electron-photon coupling, $S$ is the area
of cavity, $n_e$ is the density of electrons in QW,
$$d_{ij}=e\langle\psi_i(z)|z|\psi_j(z)\rangle$$ is the matrix
element of dipole moment, $\Delta$ is the intersubband gap,
$\epsilon_0$ is the vacuum permittivity, $\epsilon$ is the
dielectric permittivity, $L$ is the length of the cavity,
\begin{equation}\label{omega}
\omega_0(q)=\frac{c}{n}\sqrt{q^2+\left(\frac{\pi}{L}\right)^2}
\end{equation}
is the frequency of cavity photon with the wave vector
$\mathbf{q}=(q_x,q_y)$, $n$ is the refractive index of the cavity,
and $q_0$ is the resonance wave vector of cavity photons, which is
defined by the equality $\hbar\omega_0(q_0)=\Delta$ and marked at
the insert in Fig.~\ref{fig:fig3}b
\cite{Kavokin,KavokinBook,Deveaud, Sanvitto}.

In the absence of a magnetic field, the energy spectrum of
electron subbands in QW consists of a set of equidistant
parabolas,
$\varepsilon_n(\mathbf{k})=\varepsilon_n+\hbar^2\mathbf{k}^2/2m^\ast$,
where $m^\ast$ is the effective mass of electron and
$\mathbf{k}=(k_x,k_y)$ is the electron wave vector \cite{Ando,
Nag}. Therefore, the resonant energy of optical transitions is
$\Delta=\varepsilon_2-\varepsilon_1$ at any electron wave vector
(see Fig.~\ref{fig:fig11}a). As a consequence, all optical
transitions occur at the same energy, providing the emergence of a
narrow absorption line. Its linewidth, $\Gamma_0$, is defined
principally by the quality factor of the optical cavity with
additional contributions coming from the account of the finite
value of photon wave vector \cite{Kyriienko} and terms of
spin-orbit interaction \cite{KyriienkoSO}. Correspondingly, the
regime of strong light-matter coupling turns into the regime of
weak light-matter coupling under the condition $g_R=\Gamma_0$,
where the light-matter coupling constant, $g_R$, is defined by
Eqs.~(\ref{gR})--(\ref{eq:g}). Since both parameters, $g_R$ and
$\Gamma_0$, depend on fixed cavity properties, the tunable
switching of polaritonic devices between the weak and strong
coupling regimes is technically difficult. Let us demonstrate that
this problem can be easily solved with applying a magnetic field
to QW.

In the presence of an in-plane magnetic field, $\mathbf{B}=(0,B)$,
the energy spectrum of electron subbands in the QW reads
\begin{eqnarray}
\varepsilon_{n}(\mathbf{k})=\varepsilon_n+\frac{\hbar^2{\bf
k}^2}{2m^*}+\frac{\hbar k_x|d_{nn}|B}{m^*}, \label{eq:energyel}
\end{eqnarray}
where the last term describes the diamagnetic shift of electron
energy \cite{Ando, Nag}. Due to this term, the subbands in
Fig.~{\ref{fig:fig11}}b are shifted relative to each other by the
wave vector
\begin{equation}\label{dk}
\Delta k=\frac{|d_{11}-d_{22}|B}{\hbar}.
\end{equation}
It should be stressed that this shift arises from the inequality
$d_{11}\neq d_{22}$ and, therefore, takes place only in QWs with
an asymmetric confinig potential, $U(z)$. It is seen in
Fig.~\ref{fig:fig11}b that the shift (\ref{dk}) results in
different energies of intersubband optical transitions for
different wave vectors $k_x$. As a consequence, the linewidth of
the intersubband transitions achieves the additional
magnetic-induced broadening, $\Gamma_B$, which can be
approximately written as
\begin{eqnarray}
\Gamma_B\approx\frac{2\hbar k_F B|d_{11}-d_{22}|}{m^\ast},
\label{eq:B}
\end{eqnarray}
where $k_F=\sqrt{2\pi n_e}$ stands for the Fermi wave vector of
electrons. Therefore, the conversion of the regime of strong
light-matter coupling into the regime of weak light-matter
coupling is defined by the condition $g_R=\Gamma_0+\Gamma_B$.
Since the broadening (\ref{eq:B}) grows with increasing magnetic
field, this condition is satisfied at the critical magnetic field
\begin{equation}
B_{0}\approx\frac{m^*(g_R-\Gamma_0)}{2\hbar k_F |d_{11}-d_{22}|}.
\label{eq:Bcr}
\end{equation}
As a result, the range of weak magnetic fields, $B<B_0$,
corresponds to the regime of strong light-matter coupling, whereas
the range of strong magnetic fields, $B\geq B_0$, corresponds to
the regime of weak light-matter coupling. One can expect that the
field-induced switching between these two different regimes will
be crucial for physical properties of intersubband polaritons. To
concretize this, let us discuss the effect of magnetic field on
polaritonic dispersion.

\begin{figure}[h!]
\includegraphics[width=0.5\textwidth]{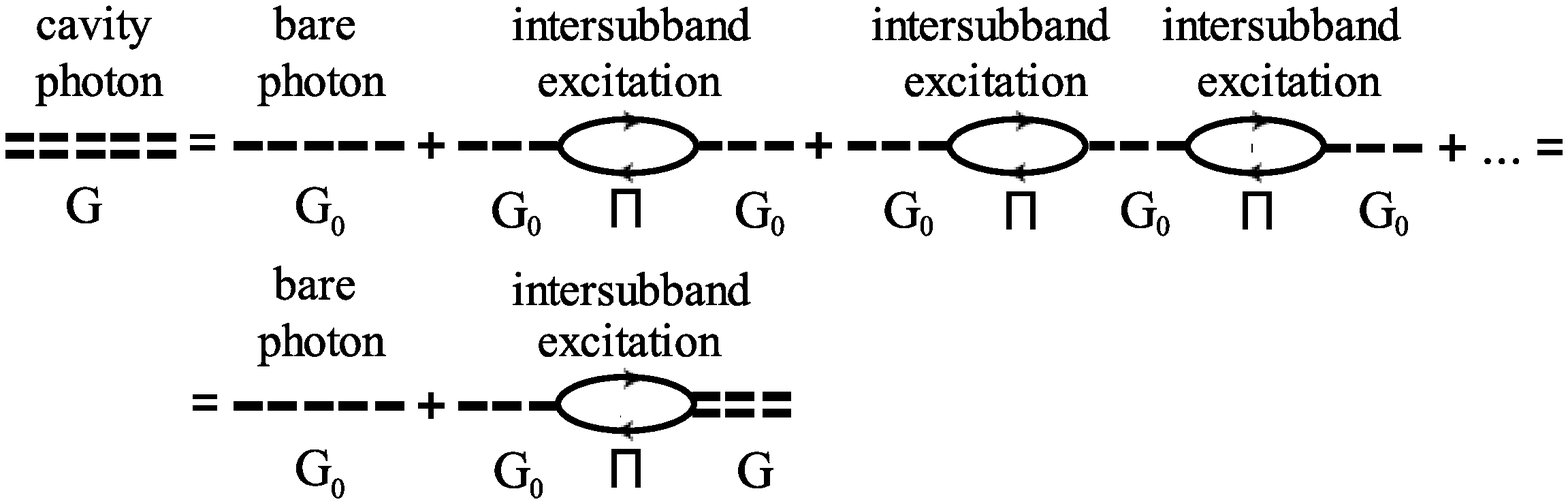}
\caption{Dyson equation for intersubband polaritons in QW.}
\label{fig:fig2}
\end{figure}

\begin{figure}[h!]
\includegraphics[width=0.47\textwidth]{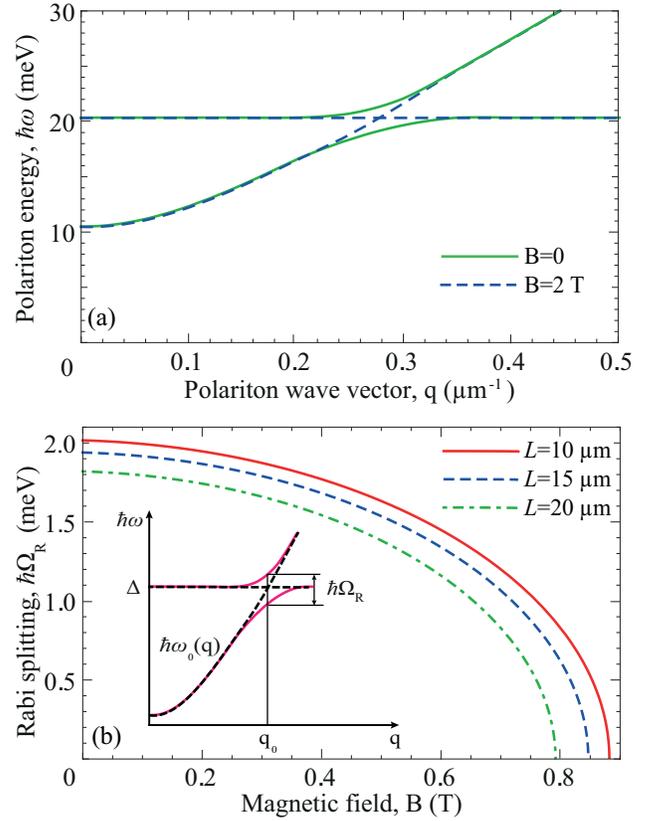}
\caption{Dispersion of intersubband polaritons for GaAs-based QW
with the electron density $n_e=10^{11}$ cm$^{-2}$ and the natural
broadening of cavity photons, $\Gamma_0=0.1$ meV: (a) the
polariton energy, $\hbar\omega$, as a function of polariton wave
vector, $q$, for the cavity length $L=10$~$\mu$m and different
magnetic fields, $B$; (b) the Rabi splitting, $\hbar\Omega_R$, as
a function of magnetic field, $B$, for different cavity lengths,
$L$. The insert shows genesis of the polaritonic branches (solid
lines) which originate from mixing electron states and cavity
photons near the resonant point, $\Delta=\hbar\omega_0(q_0)$,
where $\Delta$ is the intersubband gap and $\hbar\omega_0(q)$ is
the energy of cavity photons.} \label{fig:fig3}
\end{figure}

To calculate the dispersion of intersubband polaritons, we will
approximate the confining potential of QW by the triangular well,
\begin{equation}\label{U}
U(z)=\left\{\begin{array}{rl} \infty,
&z\leq0\\
eEz, &z>0
\end{array}\right.,
\end{equation}
where $E\approx en_e\epsilon\epsilon_0$ is the effective electric
field in QW \cite{Ando}. Applying the approach based on Green's
functions technique \cite{Kyriienko, KyriienkoSO} to the system in
question, the formation of an intersubband polariton in QW can be
described by the infinite sum of the diagrams corresponding to the
absorptions and emissions of the cavity photons, which is pictured
schematically in the first line of Fig.~\ref{fig:fig2}. This
series can be reduced to the Dyson equation for the renormalized
Green's function of cavity photon, $G(\omega,\mathbf{q})$, which
corresponds to the second line of Fig. \ref{fig:fig2}. The
solution of this equation reads
\begin{eqnarray}\label{G}
G(\omega,\mathbf{q})=\frac{G_0(\omega,\mathbf{q})}{1-g^2(q)G_0(\omega,
\mathbf{q})\Pi(\omega,\mathbf{q})},
\end{eqnarray}
where
\begin{eqnarray}
G_0(\omega,q)=\frac{2\hbar\omega_0(q)}{\hbar^2\omega^2-\hbar^2\omega^2_0(q)+2i\Gamma_0\hbar\omega_0(q)}
\end{eqnarray}
is the Green's function of bare photon,
\begin{eqnarray}
\Pi(\omega,q)&=&-2i\sum_{\mathbf{k}}\int\frac{d\nu}{2\pi}G(\nu+\omega,{\bf k+q})G(\nu,{\bf k})=\nonumber\\
&=&2\sum_{\mathbf{k}}\frac{1}{\hbar\omega+\varepsilon_{1}(\mathbf
{k})-\varepsilon_{2}(\mathbf{k+q})+i\Gamma_0} \label{eq:poloper}
\end{eqnarray}
stands for the polarization operator of intersubband transition,
where the summation is over filled electron states $\bf{k}$ under
the Fermi energy, $\varepsilon_F$. The sought polaritonic
dispersion, $\omega(\mathbf{q})$, corresponds to poles of the
Green's function (\ref{G}) and can be easily found numerically
from the equation
\begin{eqnarray}
1-g^2(q)G_0(\omega,\mathbf{q})\Pi(\omega,\mathbf{q})=0.
\label{eq:transc}
\end{eqnarray}
As expected, Eqs.~(\ref{G})--(\ref{eq:transc}) at $B=0$ turn into
the known expressions describing intersubband polaritons in the
absence of magnetic field \cite{Kyriienko}.
\begin{figure}[h!]
\includegraphics[width=0.47\textwidth]{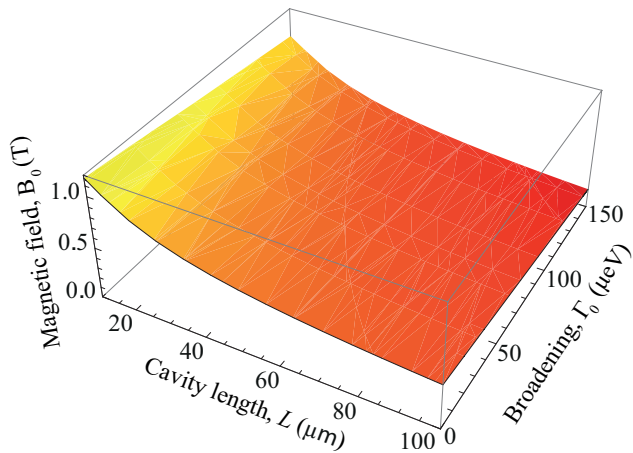}
\caption{Dependence of the critical magnetic field, $B_0$, on the
nature broadening of cavity photons, $\Gamma_0$, and the cavity
length, $L$, for GaAs-based QW with the electron density
$n_e=10^{11}$ cm$^{-2}$.} \label{fig:fig5}
\end{figure}

Solving the transcendental equation (\ref{eq:transc}),  we arrive
at the polariton dispersion which is plotted in
Fig.~\ref{fig:fig3}. In the absence of magnetic field, the regime
of strong light-matter coupling is realized. In this case, the
characteristic feature of the dispersion is the energy gap between
the two polariton branches. This gap, $\hbar\Omega_R$, takes place
at the resonant photon wave vector, $q_0$, which corresponds to
the intersection of the intersubband resonant energy, $\Delta$,
and the cavity photon dispersion, $\hbar\omega_0(q)$ (see the
insert in Fig.~\ref{fig:fig3}b). Since the gap arises from the
intersubband Rabi oscillations of electrons, it describes the Rabi
splitting of the polariton branches and depends on the Rabi
frequency, $\Omega_R$. If a magnetic field is strong enough
($B\geq B_0$), the regime of light-matter coupling is switched to
weak one. In this case, the intersubband Rabi oscillations of
electrons are broken and the Rabi splitting vanishes. The critical
magnetic field, $B_0$, which corresponds to the transition between
these two regimes, depends on parameters of the cavity and QW (see
Fig.~\ref{fig:fig5}).

Summarizing the aforesaid, we can conclude that a magnetic field
can switch a polariton system in a quantum well between the
regimes of strong light-matter coupling and weak light-matter
coupling. As a consequence, the field crucially changes the
polariton dispersion. Namely, in the regime of strong coupling the
polaritonic branches are split because of intersubband Rabi
oscillations, whereas in the regime of weak coupling this Rabi
splitting vanishes. Since all physical properties of polaritons
depends on the dispersion, magnetic field can serve as an
effective tool to control polaritonic devices.

The work was partially supported by FP7 IRSES project QOCaN, FP7
ITN project NOTEDEV, Rannis project BOFEHYSS, Rannis Excellence
Project ``2D transport in the regime of strong light-matter
coupling'', Singaporean Ministry of Education under AcRF Tier 2
grant MOE2015-T2-1-055, RFBR project 14-02-00033 and the Russian
Ministry of Education and Science. I.A.S. thanks the support of
the 5-100 program of the Russian Federal Government. We thank O.
Kyriienko for valuable discussions.

\end{document}